\documentclass[lettersize,journal]{IEEEtran}
\usepackage{amsmath,amsfonts}
\usepackage{algorithmic}
\usepackage{array}
\usepackage[caption=false,font=normalsize,labelfont=sf,textfont=sf]{subfig}
\usepackage{textcomp}
\usepackage{stfloats}
\usepackage{url}
\usepackage{verbatim}
\usepackage{graphicx}
\def\BibTeX{{\rm B\kern-.05em{\sc i\kern-.025em b}\kern-.08em
    T\kern-.1667em\lower.7ex\hbox{E}\kern-.125emX}}
\usepackage{balance}
\begin{document}
\title{A Survey on Federated Learning in Intelligent Transportation Systems}

\author{Rongqing~Zhang,
Hanqiu~Wang, 
Bing~Li, Xiang~Cheng, and~Liuqing~Yang
\thanks{R. Zhang, H. Wang and, B. Li are with the School of Software Engineering, Tongji University, Shanghai 201804, China (e-mail: rongqingz@tongji.edu.cn; 2031562@tongji.edu.cn; lizi@tongji.edu.cn).}
\thanks{X. Cheng is with the School of Electronics, Peking University, Beijing 100871, China (e-mail: xiangcheng@pku.edu.cn).}
\thanks{L. Yang is with the Internet of Things Thrust and Intelligent Transportation Thrust, Hong Kong University of Science and Technology (Guangzhou),  Guangzhou 511458, China, and also with the Department of Electronic and  Computer Engineering, Hong Kong University of Science and Technology, Hong Kong SAR 5999077, China (e-mail: lqyang@ust.hk).}
}

\maketitle

\begin{abstract}

The development of Intelligent Transportation System (ITS) has brought about comprehensive urban traffic information that not only provides convenience to urban residents in their daily lives but also enhances the efficiency of urban road usage, leading to a more harmonious and sustainable urban life. Typical scenarios in ITS mainly include traffic flow prediction, traffic target recognition, and vehicular edge computing. However, most current ITS applications rely on a centralized training approach where users upload source data to a cloud server with high computing power for management and centralized training. This approach has limitations such as poor real-time performance, data silos, and difficulty in guaranteeing data privacy. To address these limitations,  federated learning (FL) has been proposed as a promising solution. In this paper,  we present a comprehensive review of the application of FL in ITS, with a particular focus on three key scenarios: traffic flow prediction, traffic target recognition, and vehicular edge computing. For each scenario, we provide an in-depth analysis of its key characteristics, current challenges, and specific manners in which FL is leveraged. Moreover, we discuss the benefits that FL can offer as a potential solution to the limitations of the centralized training approach currently used in ITS applications.

\end{abstract}


\section{Introduction}

\begin{figure}[htbp]
	\centering
	\includegraphics[width=3.3in]{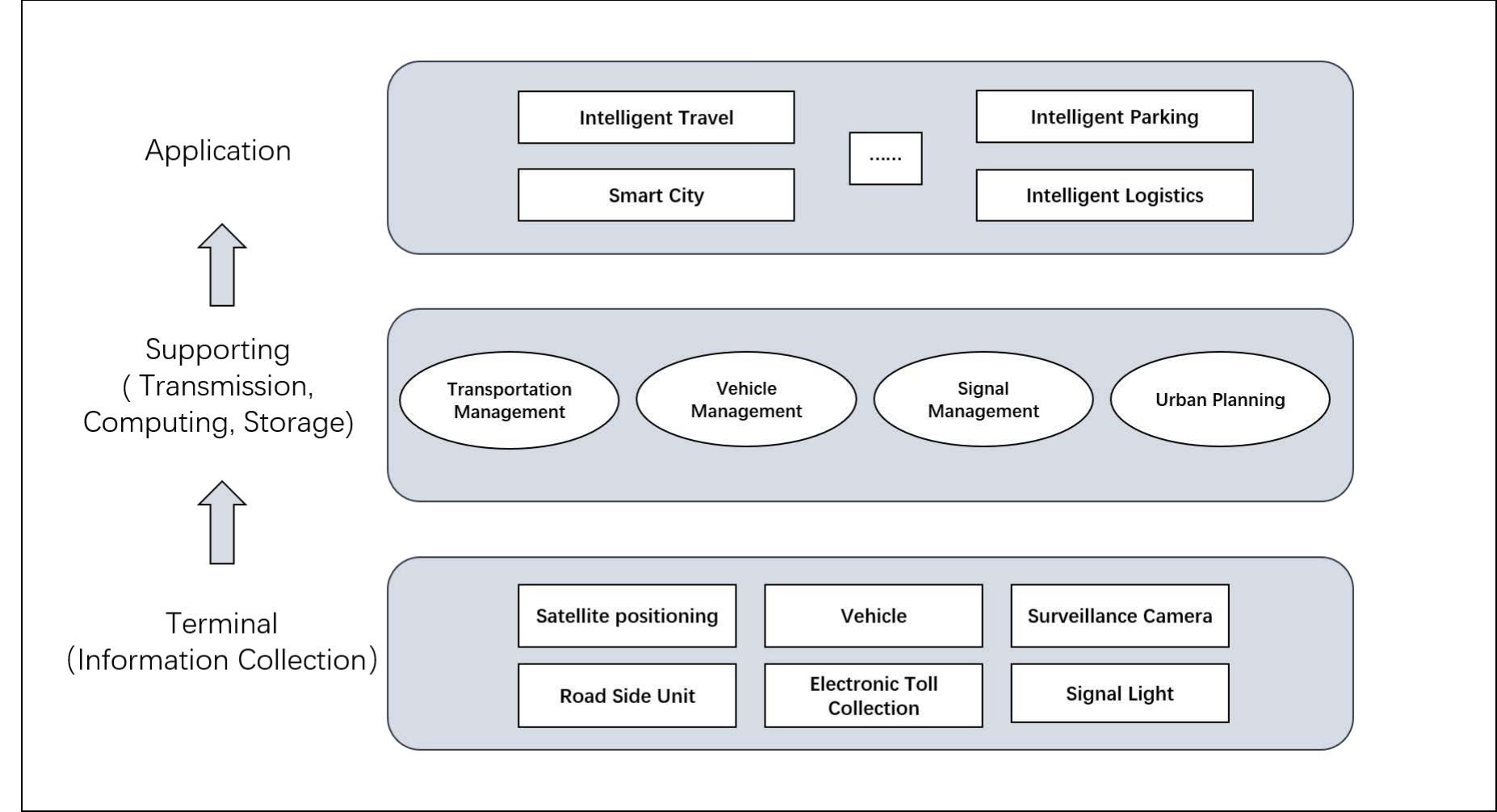}
	\caption{Framework of a typical ITS.} \label{Fig1.1_ITS}
	\vspace{-6mm}
\end{figure}

In recent years, the accelerated urbanization process has led to a marked increase in vehicular traffic, necessitating the development of an efficient and effective traffic management system. The Intelligent Transportation System (ITS), initiated in the early 1970s, is a traffic management system aimed at enhancing transportation efficiency
 \cite{Zhang2011}. ITS capitalizes on a robust infrastructure and incorporates advanced technologies to improve transportation system performance. Its foremost goal  is to integrate diverse road information comprehensively  and to fully leverage the power of the internet, cloud computing, artificial intelligence, and other advanced technologies within the transportation field. This integration will enable the achievement of comprehensive information perception and the ability to make scientifically informed decisions \cite{Zhu2018}. A conceptual overview of the typical ITS framework is depicted in Fig.~\ref{Fig1.1_ITS}.

Comprehensive urban traffic information is an essential component of modern urban transportation management, offering benefits to both urban residents and the environment. The deployment of ITS can  enhance the utilization of transportation resources and improve traffic efficiency by leveraging advanced  technologies such as artificial intelligence and machine learning. These technologies equip drivers with essential information, facilitating informed decision-making in urban road navigation. Moreover, the deployment of ITS has the potential to  mitigate  environmental pollution by reducing  congestion and the associated carbon emissions. ITS provides real-time traffic information, which helps drivers avoid congestion and take alternative routes, thereby reducing the number of vehicles on roads  and decreasing carbon emissions. The reduction in emissions contributes to the development of a sustainable and harmonious urban environment. As such, ITS has  emerged as  a representative application in the field of Internet of Things (IoT). The main scenarios in which ITS are commonly deployed include traffic flow prediction, traffic target recognition, and vehicular edge computing.


Traffic flow prediction is crucial  in  ITS for effective  traffic management, control, and planning \cite{Lv2014}. With the rapid growth of urbanization and vehicular technology in recent years, the number of vehicles in urban areas has increased significantly, leading to traffic congestion and related challenges. Traffic flow prediction estimates  future traffic flow by analyzing historical traffic data patterns. Accurate traffic forecasting information can provide crucial guidance for urban road network planning and enable the rapid real-time prediction of complex traffic scenarios, thus alleviating traffic congestion. Furthermore, comprehensive traffic information can help residents make more informed travel plans, provide more efficient route navigation, and create a safe and comfortable public  travel environment, thereby improving the safety and efficiency of the transportation system.


Traffic target recognition is also a crucial application in ITS that encompasses the detection and identification of both road targets and in-vehicle targets through  technologies such as image recognition. With the recent advancements in autonomous driving technology, target recognition in traffic scenes has gained increased attention. Target recognition involves identifying road targets such as traffic signs, obstacles, and pedestrians, as well as in-vehicle targets including drivers, passengers, and objects. Comprehensive traffic information provided by accurate target recognition can benefit both city management and residents by aiding in transportation planning and improving the efficiency and safety of the transportation system.


Vehicular edge computing is a promising strategy for the efficient execution of application tasks, facilitating the offloading of tasks from smart vehicles to the edge for processing. The offloading employs  a wireless link between the smart vehicle and the Road Side Unit (RSU) to transfer task data and obtain processing results. With the increasing volume of urban traffic data and the demand for real-time information processing, traditional methods, which transfer all data back to the cloud computing center, often lead to bandwidth wastage and increased latency. Vehicular edge computing enhances computational efficiency and leverages edge devices to achieve efficient utilization of resources. By analyzing and processing data in real-time at the edge, it offers immediate guidance based on current road conditions and available resources. The utilization  of edge computing provides several benefits, including reduced latency, increased bandwidth utilization, and lower energy consumption.


The centralized training mode is widely used in current  ITS applications, as illustrated in Fig.~\ref{Fig1.2_Centralized}. This approach involves uploading source data to a cloud server with high computing power for management and centralized training. However, it has several limitations:

\begin{figure}[htbp]
	\centering
	\includegraphics[width=3.3in]{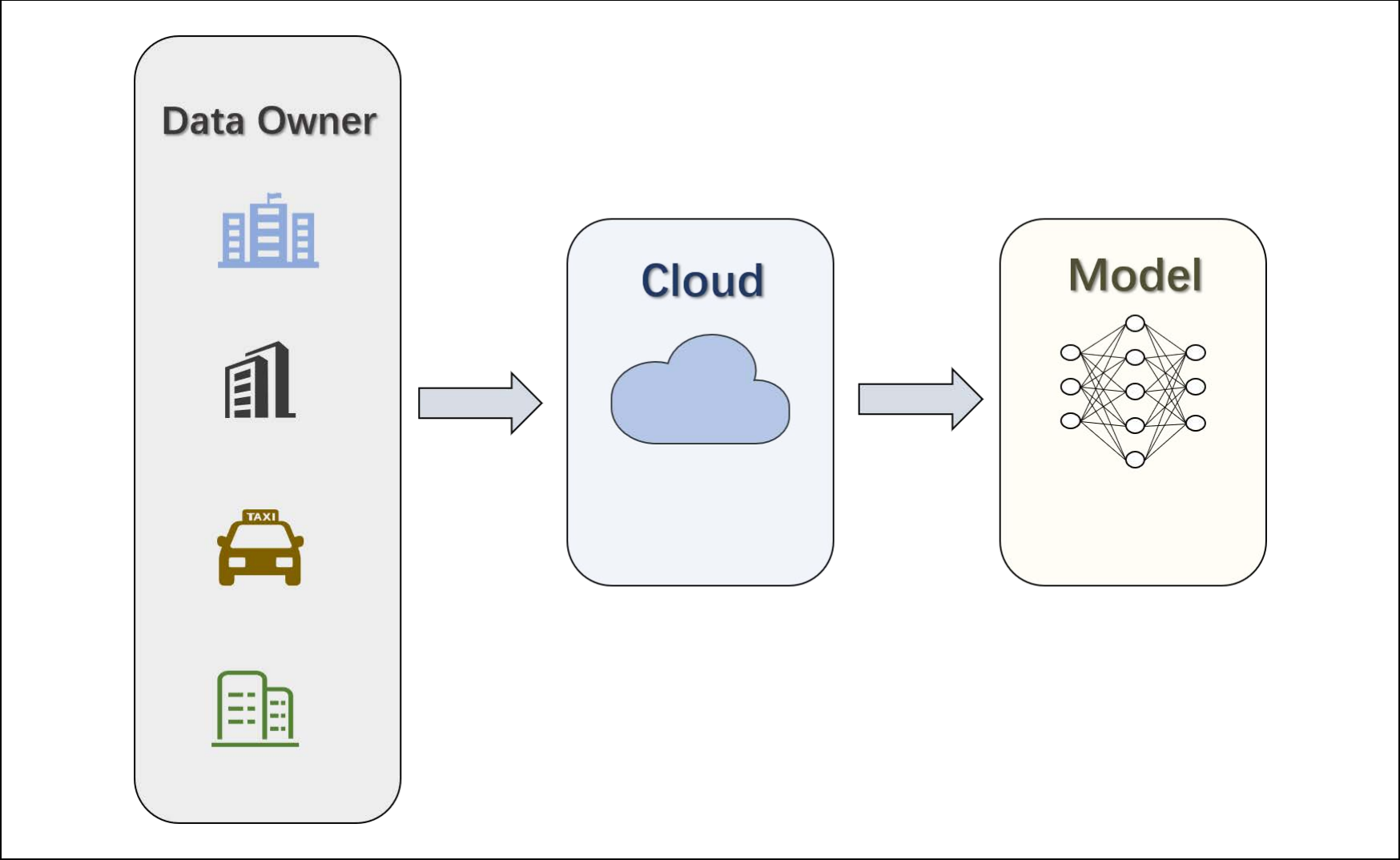}
	\caption{Centralized training.} \label{Fig1.2_Centralized}
	\vspace{-3mm}
\end{figure}

\begin{figure}[htbp]
	\centering
	\includegraphics[width=3.3in]{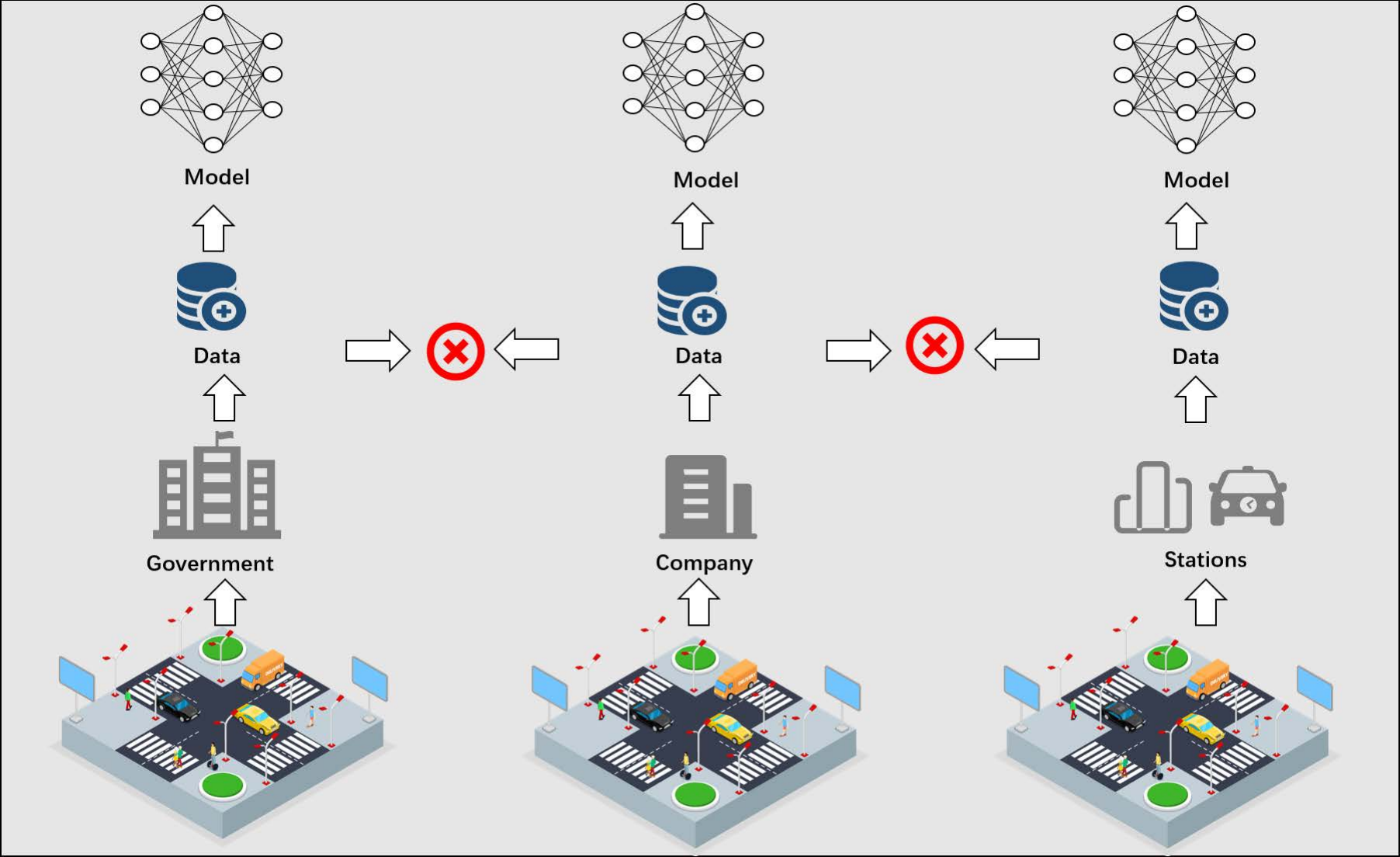}
	\caption{Data silos.} 
	\label{Fig1.3_Data_silos}
	\vspace{-2mm}
\end{figure}

\begin{itemize}
%
	\item Data privacy. The centralized training approach aggregates global data and collects all user data on a centralized server for unified management, without considering  the privacy information contained in different organizational data sets. As a result, the centralized training approach can lead to uncontrolled data flows  and sensitive data leakage. For instance, in traffic flow prediction scenarios, different data owners may have access to different types of data, such as cab companies owning users' trajectory data and governments owning road data, which may contain users' private information. Uploading all the data to a central server may result in potential privacy leakage, leading to significant privacy and security concerns among  users. Therefore, the centralized training approach is limited in terms of ensuring user security and protecting data privacy.
	
	
	\item Poor real-time performance. The massive data generated in ITS requires a large amount of computing resources, which poses a challenge for centralized servers to meet the real-time performance requirements of users. In addition, the  delay caused by transmitting data to a remote server for processing and then returning the results to the user also reduces the speed of response. Such  delay may have a significant impact on time-sensitive applications, such as real-time traffic flow prediction and autonomous driving. 
	
	
	\item Data silos. Due to network security isolation and industry privacy concerns, data barriers exist between different organizations, departments, and systems, resulting in data silos that prevent secure data sharing. This leads to the fact that some of the valuable data cannot be used to train an effective model, resulting in a significant bottleneck for many applications, as illustrated in Fig.~\ref{Fig1.3_Data_silos}. 
\end{itemize}

The  limitations have motivated the development of decentralized training approaches that can address the  challenges and enable more efficient and secure data processing in ITS. Hence, in view of the aforementioned limitations of centralized training, federated learning (FL) is proposed as a potential solution. This paper provides a comprehensive review of the application of FL in ITS, concentrating  on three principal  scenarios: traffic flow prediction, traffic target recognition, and vehicular edge computing. For each scenario, we provide an overview of its key characteristics and current challenges, outline the specific ways in which FL is leveraged, and elaborate on the benefits that FL can bring as a potential solution.


\section{Federated Learning}


As data privacy concerns intensify, sharing data between different organizations becomes more challenging. Given that data frequently contains highly sensitive private information,  owners stringently restrict its sharing. Additionally, with the explosion of data volume, the computation and communication costs of centralized data processing can be unaffordable. As a result, research has suggested FL as a method  to protect machine learning privacy \cite{Yang2019, McMahan2017}.

\begin{figure}[htbp]
\vspace{-3mm}
	\centering
	\includegraphics[width=3.3in]{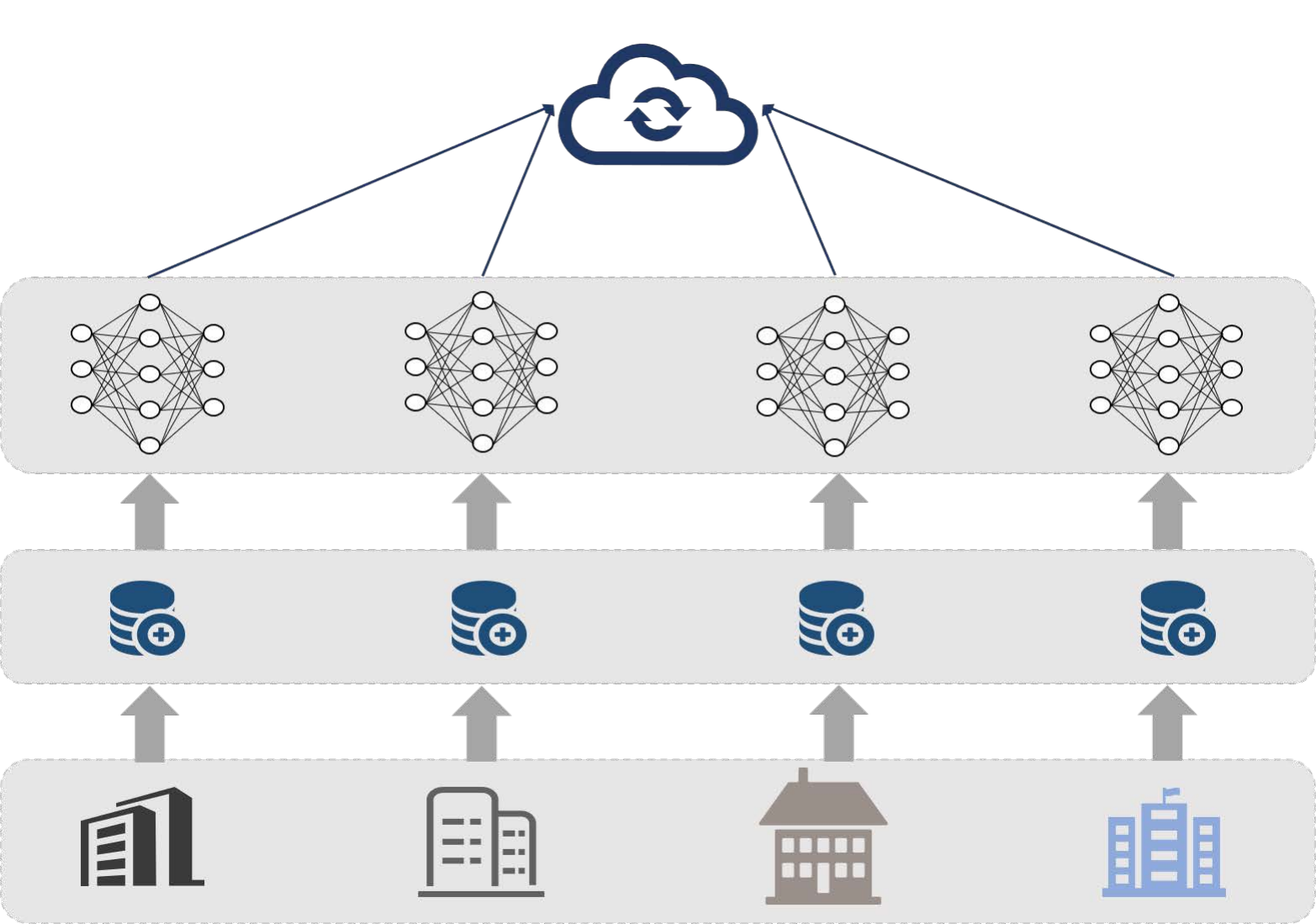}
	\caption{FL framework.} 
	\label{Fig2.1_FL}
	\vspace{-2mm}
\end{figure}

FL is a novel training method within the distributed model training framework. As illustrated  in Fig.~\ref{Fig2.1_FL}, this method consists of multiple participating clients, each possessing  a subset of local data for training, coupled with a central server responsible for aggregating the clients' model updates. FL enhances data isolation during model training, protects data privacy, and enables efficient collaborative learning among multiple participants. It  allows both the storage of data and the model training phases to occur locally, thereby promoting collaborative training of globally optimal models. Through exclusive reliance on central server interactions for model updates, FL has shown great potential in  improving the efficiency and security of machine learning within  ITS applications, leading to extensive research and development in recent years\cite{Bonawitz2019}.

\begin{figure}[htbp]
	\centering
	\includegraphics[width=3.3in]{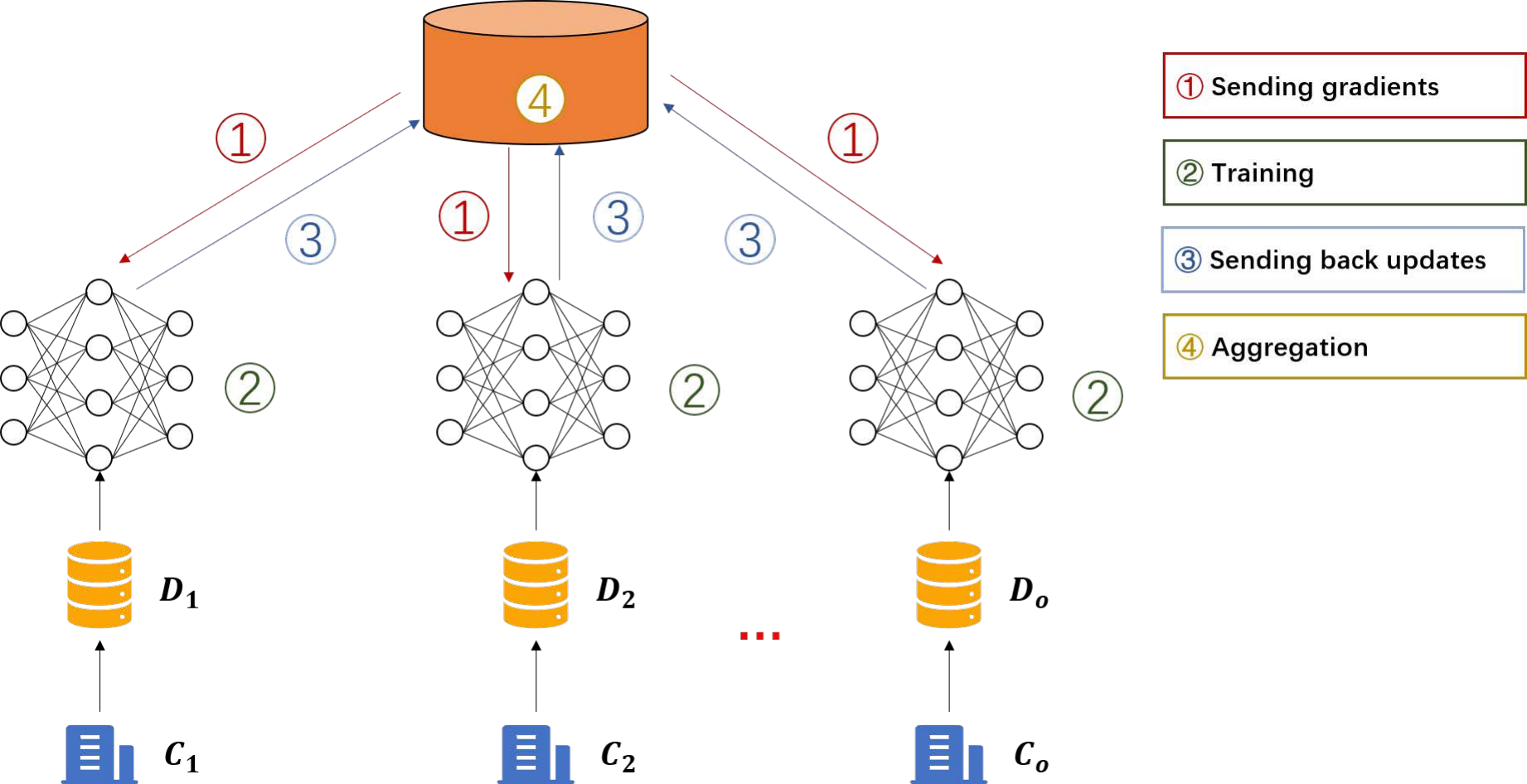}
	\caption{FL steps.} 
	\label{Fig2.2_Fl_steps}
\end{figure}

In FL,  model training is performed locally, and each organization uploads the encrypted model parameters to the cloud after the completion of local model training. The cloud then aggregates the parameters to obtain the updated  model parameters, which are subsequently redistributed  to  each organization for model updating. Recent studies have shown that FL provides a trade-off between model performance and privacy protection \cite{Lu2019}. This is because, in FL, data remains locally with the organizations, and only model parameters are shared with the cloud. This approach can reduce the risk of sensitive data exposure  during model training. Moreover, the distributed structure  of FL allows organizations to collaborate on model development without the need for a centralized data repository. The FL methodology entails four primary stages, which are illustrated in Fig.~\ref{Fig2.2_Fl_steps}.

\begin{enumerate}

	\item In FL, the initiation phase involves establishing a global model, which is randomly initialized by the cloud server in the first round of training. This model is then transmitted to each participating organization, along with its weights. In subsequent rounds, the cloud server sends the current global model weights to each organization.

	
	\item During the FL process, participating organizations utilize their own local data to train the model and iteratively update the model's weights.
	
	
	\item Each participating organization sends the updated model weights to the cloud server without sharing their local data, thus preserving data privacy and confidentiality.
	
	
	\item Following the receipt of updated model weights from each participating organization, the cloud server aggregates the model weights and updates the global model. Subsequently, the cloud sends the updated model weights to the participating organizations, and this iterative process is repeated until the global model converges.
	
\end{enumerate}

%
%

The aforementioned steps highlight several notable advantages of FL. Firstly, FL enables data isolation by  ensuring the confidentiality of local data  throughout the training process, aligning with the requirements  for user data privacy protection in modern data-driven applications. Secondly, FL supports multi-party collaboration, where each participant contributes their training data to build a shared global machine learning model, thereby promoting equality among all parties involved in the process. Finally, FL allows each organization to maintain  independence while simultaneously exchanging model weights, which enhances the global model's efficiency. The combination of these  advantages makes FL a promising approach for developing collaborative machine learning solutions within  distributed environments.

Accordingly, we introduce FL into the within  of ITS and provide an overview of current related research.

\section{Federated Learning in Traffic Flow Prediction}

\subsection{Traffic Flow Prediction Scenario}

\begin{figure}[htbp]
	\centering
	\includegraphics[width=3.3in]{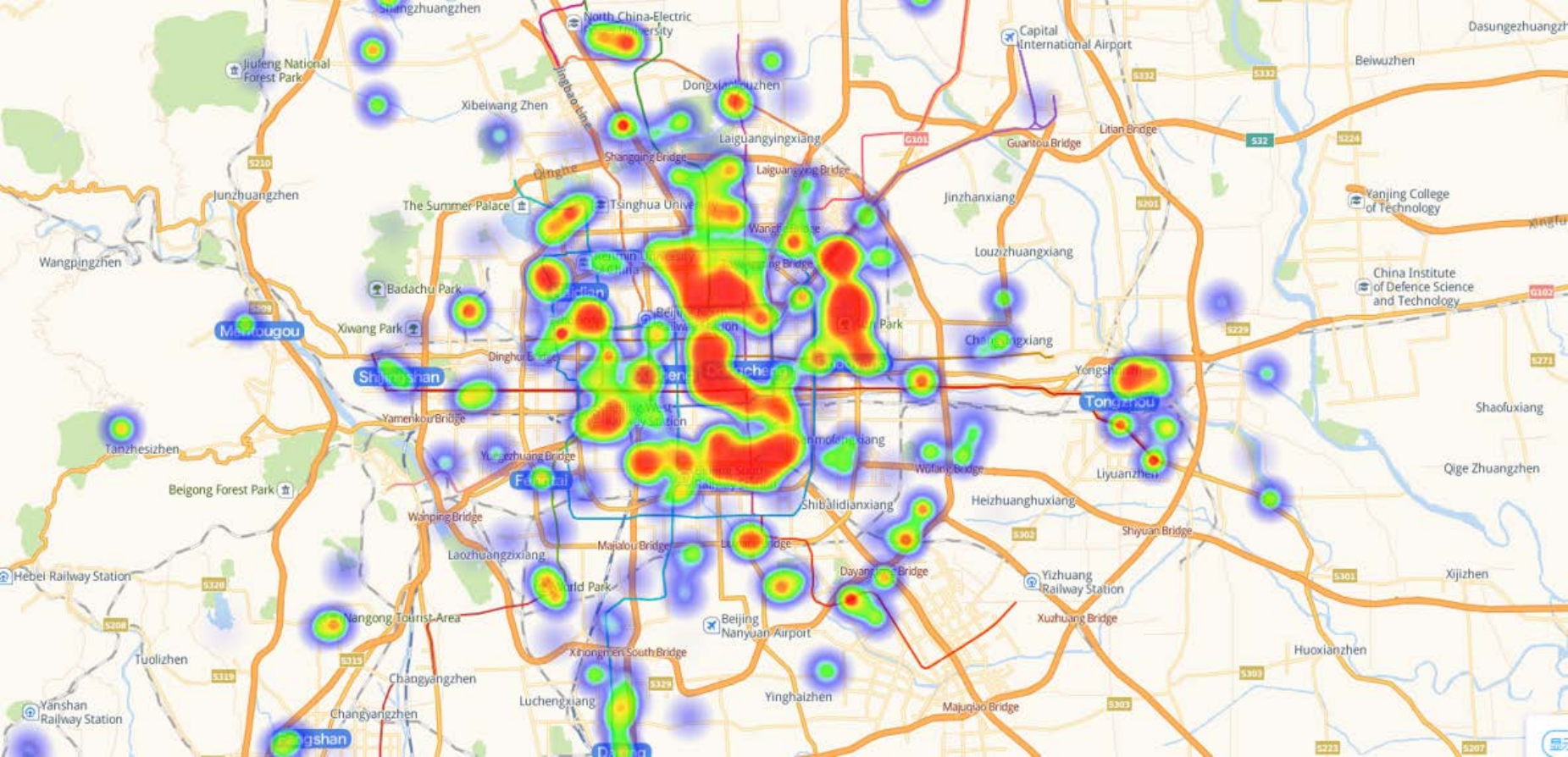}
	\caption{A heatmap of Beijing, China} 
	\label{Fig3.1_Heatmap}
	\vspace{-3mm}
\end{figure}

%
%
%

Traffic flow prediction is a critical issue within  the field of ITS. It involves the estimation of future traffic volume in  a specific  area by analyzing the patterns of traffic changes based on historical  data. Fig.~\ref{Fig3.1_Heatmap} illustrates  a heatmap of  traffic conditions in Beijing, China, serving as a visual representation of the city's current traffic flow. The provision of accurate traffic forecasting  is essential for facilitating timely and reliable traffic condition predictions, enabling individuals to make travel plans that optimize travel time and enhance the safety and efficiency of the transport system \cite{Tedjopurnomo2020}. In general, traffic flow prediction scenarios exhibit four key characteristics:

\begin{enumerate}
	
	\item High privacy. The challenge  of maintaining  high privacy is paramount  in the context of traffic flow prediction. This is because GPS track data used for this purpose often contains sensitive information regarding users' daily behaviors, which cannot be shared freely.

	
	\item Rich types of data. The various scenarios associated with traffic flow prediction are characterized by diverse data types, including weather features, spatio-temporal features, POI (Point of Interest) semantic features, connection relationships, among others. Such rich data can enable a multifaceted portrayal of traffic flow changes.
	

	\item High real-time requirements. Short-term traffic forecasting  relies on real-time information, emphasizing the criticality of prompt  feedback to the road and  timely adjustments to vehicle decisions. Consequently, addressing this challenge forms a central focus in the traffic prediction problem.
	
	
	\item Sensitive data structure features. The data used for traffic flow prediction is characterized by sensitive data structure features, containing not only intersection feature data but also topology information. This presents a potential privacy risk when data providers in the ITS domain, such as government agencies and companies, exchange data during cooperation and sharing activities. The topological information may contain sensitive data, thus requiring careful handling and management to prevent unauthorized data access and privacy breaches.
\end{enumerate}

Numerous studies have been conducted to address the challenges posed by the characteristics of traffic flow prediction, as shown in Table \ref{Tab:Federated Learning in Traffic Flow Prediction}.

\subsection{Federated Learning in Traffic Flow Prediction}


The majority of current traffic prediction research is focused on exploiting the spatio-temporal correlation of urban traffic, and extracting the unique features and inter-region correlations within traffic data. Among the popular models for traffic flow prediction, the Graph Convolutional Neural Network (GCN) has emerged as a promising approach due to its ability to incorporate the natural graph structure properties of traffic data and integrate spatio-temporal features \cite{Guo2019,Yu2018,Wang2022}. However, during the model training process, the GCN is associated with several issues, such as high data privacy risks, time-consuming training, and high communication costs, which pose significant challenges to the development of traffic flow prediction research.

\begin{table*}[htbp]
	\renewcommand{\arraystretch}{2.5}
	\caption{Federated Learning in Traffic Flow Prediction. } 
	\centering
	\begin{tabular}{|p{4cm}|p{2cm}|p{4cm}|p{4cm}|}
		\hline 
		Applications & Reference & Main Ideas & Advantages \\
		\hline
		Traffic Flow Prediction &
		\cite{Xia2022} &
		Replace the global road network model by horizontal local road network. &
		Reduce computing and communication costs while preventing data privacy exposure.\\ 
		\cline{2-4}
		& \cite{Zhang2021}&
		Propose a differential adjacency matrix protection method. &
		Protect the topological information of the traffic network.\\
		\cline{2-4}
		& \cite{Yuan2022}&
		Develope the Federated Graph Attention layer to share parameters based on VFL. & 
		Address the problem of participants with different spatial characteristics.\\
		\cline{2-4}
		& \cite{Qi2021,Meese2022}&
		Design a blockchain-based FL framework and replace the central server by a set of trusted consensus nodes. &
		Address single point failure of server and malicious vehicles.  \\
		\cline{2-4}
		&\cite{Akallouch2022,Deng2022} & 
		Encrypt model parameters.& 
		Mitigate the inference attacks. \\
		\cline{2-4}
		& \cite{Liu2020}& 
		Improve the federated averaging algorithm with clustering organizations based on latitude and longitude.& 
		Enhance the scalability of FL and reduce communication overhead. \\
		\cline{2-4}
		& \cite{Zhang2022} &
		Propose a client clustering approach based on local model similarity. & 
		Reduce communication costs on transmission of model updates.
		\\
		\cline{2-4}
		& \cite{Yuan2022a} &
		Design an asynchronous algorithm for model parameter uploading and downloading decisions. & 
		Enhance the communication efficiency of FL.\\
		\cline{2-4}
		\hline
		Travel Time Estimation
		& \cite{Zhang2022a} &
		Construct a global modelas shared by all clients and a fine-tuned personalized model for each client to capture individual driving habits.& 
		Address non-IID caused by personal driving habits and the inconsistency with data among clients. \\
		\cline{2-4}
		&\cite{Zhu2022} &
		Train a customized neural network travel time estimator for each area using locally collected data. & Cross-area travel time estimation.\\
		\hline
		Crowd Flow Prediction 
		& \cite{Wang2022a} &  
		Learn spatio-temporal features from human trajectory and classify clients with similar spatio-temporal features into same cluster. &
		Improve the efficiency and accuracy of FL-based human mobility prediction.\\
		\cline{2-4}
		& \cite{Errounda2022} &
		Propose a mobility vertical federated framework that allows the learning process to be conducted over vertically partitioned data. & 
		Enable forecasting of mobility covering a joint location domain.\\
		\hline
		Route Planning
		&\cite{Zeng2021} & 
		Implemente a hierarchical clustering approach to partition the traffic data into groups, and utilized FL to train the models among all stations.& 
		Protect the privacy and reduce the communication cost.\\
		\hline
		Destination Prediction 
		&\cite{Halim2020} &
		Use blockchain instead of centralized servers in FL. & 
		Eliminate single points of failure.\\
		\hline
		Parking Management
		& \cite{Huang2021} &
		Adopt FL and LSTM to promote the collaboration in parking space estimation. &
		Protect user privacy. \\
		\hline
	\end{tabular}
	\label{Tab:Federated Learning in Traffic Flow Prediction}
\end{table*}


In response to these challenges, recent studies have proposed the use of FL in the field of traffic flow prediction. Xia \emph{et al.} \cite{Xia2022} proposed a FL integrated GCN to address the challenges of balancing accuracy and time cost while preserving privacy. By utilizing horizontal local road network FL, the global graph convolutional road network model was replaced to reduce computing and communication costs while preventing data privacy exposure. However, existing GNN-based models  employing FL in ITS  tend to overlook the topological information of the traffic network, thereby risking privacy breaches. To address this challenge, Zhang \emph{et al.} \cite{Zhang2021} proposed a differential adjacency matrix protection method. The method involved transforming the original adjacency matrix at the organization side by a Gaussian matrix to protect local topological information, and constructing the global adjacency matrix at the server side by processing the adjacency matrices from different organizations. Additionally, the FL approach in traffic flow prediction typically involves participants with the same sample space but different spatial characteristics on the traffic flow data set. To address the issue of different spatial characteristics of participants, the Vertical FL (VFL) is required. Yuan \emph{et al.} \cite{Yuan2022} developed the federated graph attention layer to capture short-term temporal information without loss of spatial information among areas by sharing parameters based on VFL.


Moreover, some studies have focused on potential security issues associated with the integration of FL in traffic flow prediction. To address such security challenges, including single point of server failure and malicious vehicles, researchers have proposed a blockchain-based FL framework \cite{Qi2021,Meese2022}. In this framework, the central server is replaced by a set of trusted consensus nodes to mitigate the risk of single point of failure. Model updates are verified by miners and stored in the blockchain, which can detect malicious vehicles and defend against potential security attacks. Another potential security risk in FL is inference attacks, where an attacker may recover data from the gradient parameters to infer the original user data by inverting the shared gradients. To mitigate this risk, some researchers have added local differential privacy on local gradients to protect the privacy information \cite{Akallouch2022}. Additionally, Deng \emph{et al.} have proposed a solution by encrypting the model parameters and using homomorphic encryption to prevent attackers from performing inference attacks \cite{Deng2022}.

%

The efficiency of communication in introducing FL in traffic flow prediction is also a critical aspect that deserves further investigation. To address the scalability challenges of FL frameworks, an improved federated averaging algorithm with random subsampling of participants was designed in \cite{Liu2020} to reduce communication overhead. A federated scheme employing 
 integrated clustering was  proposed to cluster organizations based on latitude and longitude, and the global model is eventually integrated at each clustering center. Similarly, a client clustering approach was introduced in \cite{Zhang2022}, where clients with similar model parameters are grouped into the same cluster, reducing communication costs for the transmission of model updates in the FL system. To enhance the communication efficiency of FL, an asynchronous algorithm was designed in \cite{Yuan2022a} to account for the impact of rounds on the aggregation results.


The aforementioned studies have demonstrated improved experimental results by more effectively integrating traffic prediction with FL, thus addressing the limitations of traditional traffic prediction methods.

\subsection{Federated Learning in Applications of Traffic Flow Prediction}


In addition to the challenge of predicting traffic flow, there exist numerous specific applications in the domain of traffic flow prediction that incorporate FL techniques. In this context, we offer a comprehensive explication of these applications.


Travel time estimation has emerged as a key area of research within  ITS in recent years, aiming to compute  average travel time from the origin to the destination using historical data. Accurate travel time estimation is crucial for transportation agencies and individuals to plan and optimize travel routes, reduce congestion, and improve overall traffic flow. In view of the sensitivity of user travel information, which cannot be shared due to privacy concerns, a personalized FL strategy has been developed in \cite{Zhang2022a}. This strategy leverages the FL principles to construct a global model, which is designed as an online generative model shared by all clients. This approach also incorporates a fine-tuned personalized model for each client to capture their individual driving habits and compensate for residual errors resulting from localized global model predictions. In addition, privacy and security concerns present  challenges in cross-area travel time estimation, particularly with respect to data exchange among different areas. To address these challenges, Zhu \emph{et al.} proposed a novel approach that  trains  a customized neural network travel time estimator for each area using locally collected data. The proposed approach incorporates FL for training the model, thereby enabling the exchange of information among different areas while preserving individual data privacy \cite{Zhu2022}.


In recent times, crowd flow prediction has become an increasingly critical topic. With the rapid growth of urban areas and the increasing complexity of transportation systems, there is a growing need to develop advanced technologies and methodologies to better understand and manage human mobility. The authors of \cite{Wang2022a} proposed an enhanced FL framework by incorporating clustering algorithms. This approach involves leveraging human trajectory data to extract spatio-temporal features and grouping clients with similar features into clusters. By applying clustering algorithms, the proposed approach aims to improve the efficiency and accuracy of FL-based human mobility prediction. To address the challenge of achieving predictions for the entire location domain, which can vary significantly across different organizations, a mobility vertical FL prediction framework was developed for the mobility prediction problem, as described in \cite{Errounda2022}. This framework allows for joint learning to be performed over vertically partitioned data belonging to multiple organizations. Specifically, the mobility data is vertically partitioned by location, allowing for each organization to contribute their respective data while ensuring that sensitive information is kept private.


%
%

FL has demonstrated strong performance across various applications. For instance, in a recent study by Zeng \emph{et al.} \cite{Zeng2021}, a multi-task FL framework was proposed to optimize traffic prediction models for route planning. The authors implemented a hierarchical clustering approach to partition the traffic data into groups, and utilized FL to train the models among all stations without sharing data. FL has emerged as a powerful tool for accurate location services while preserving user privacy, particularly for destination prediction tasks \cite{Halim2020}. In order to mitigate the risk of malicious users, blockchain technology can be leveraged in place of centralized servers, thereby eliminating single points of failure. FL continues to find practical applications in the domain of parking management, where privacy concerns remain a paramount consideration. Parking space information, which includes sensitive details such as the arrival and departure times of vehicles, is crucial to capturing a basic overview of daily parking requests. In a recent study by Huang \emph{et al.} \cite{Huang2021}, an FL-based approach was proposed for real-time prediction of the number of available parking spaces in a car park.


In summary, FL has demonstrated efficacy in addressing privacy concerns in traffic data and enhancing the security of the training process, applicable to both broad  traffic flow prediction and specific applications within this domain.

\section{Federated Learning in Traffic target recognition}

\subsection{Traffic Target Recognition Scenario}

%
%

\begin{figure}[htbp]
	\centering
	\includegraphics[width=3.3in]{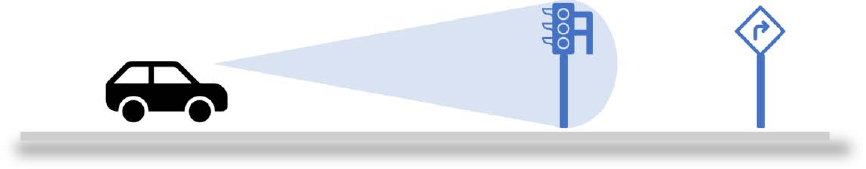}
	\caption{Road target recognition.} 
	\label{Fig4.1_Target_Recognition}
\end{figure}

\begin{figure}[htbp]
	\centering
	\includegraphics[width=3.3in]{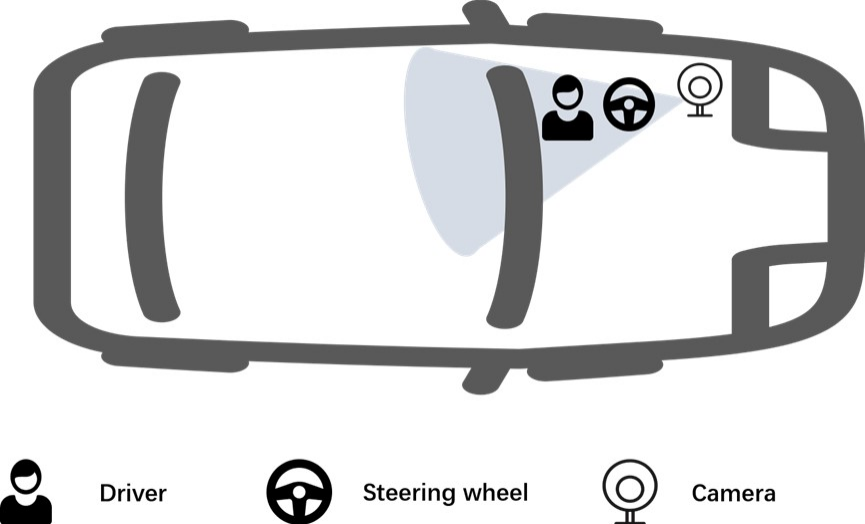}
	\caption{In-vehicle driving assistance.} 
	\label{Fig4.2_In_Vehicle_Recognition}
	\vspace{-3mm}
\end{figure}

The traffic target recognition scenario refers to the use of image recognition techniques in ITS and encompasses two main subcategories: road target recognition and in-vehicle driving assistance. Road target recognition employs computer vision techniques in automotive assisted driving systems, whereby forward-facing cameras are used to identify common traffic signs, road targets, and other relevant objects as illustrated in Fig.~\ref{Fig4.1_Target_Recognition}. The ultimate objective is to leverage this information to make informed decisions and enhance driving safety. In-vehicle driving assistance involves employing advanced technologies to provide real-time support and assistance during the driving process, such as driver emotion recognition and steering wheel angle detection as demonstrated in Fig.~\ref{Fig4.2_In_Vehicle_Recognition}. The objective  is to enhance both the convenience and safety of driving for  drivers.

The traffic road recognition scenario exhibits several distinct challenges, including:

\begin{enumerate}

	\item Lack of sufficient data. Most driving images contain private information, such as images of the driver's face, and as a result, this data cannot be shared, leading to an insufficient training data for collaborative model development.
	
	
	\item High resource consumption. The transmission of image data requires a significant amount of bandwidth and storage resources, which can be costly, especially in situations where large amounts of data need to be transmitted frequently.

	
	\item High computational load. If all data is aggregated to the center for unified processing, the computational load at the center can become overwhelming, leading to significant delays in the recognition process.
	
	\item High real-time requirements. Target recognition scenarios typically require quick recognition results to make the right decision, resulting in high computational latency requirements.
	
\end{enumerate}

Overall, the traffic target recognition scenario presents several significant challenges, including the lack of sufficient data, high resource consumption, high computational load, and high real-time requirements. Researchers are actively developing new approaches, such as FL, to address these challenges and improve the accuracy and efficiency of traffic target recognition as shown in Table \ref{Tab:Federated Learning in Traffic Target Recognition}. In the following part, we will specifically describe the application of FL in traffic target recognition scenarios.

\begin{table*}[htbp]
	\renewcommand{\arraystretch}{2.5}
	\caption{Federated Learning in Traffic Target recognition. } 
	\centering
	\begin{tabular}{|p{4cm}|p{2cm}|p{4cm}|p{4cm}|}
		\hline 
		Applications & Reference & Key contributions & Advantages 
		\\
		\hline
		\multicolumn{4}{|c|}{Road Target Recognition}
		\\
		\hline
		Pothole Detection&
		\cite{Rahman2021,Alshammari2022} &
		Apply FL to pothole detection. & 
		Address the limitations of time-consuming and prone to errors.\\
		\hline
		Road Damage Detection & 
		\cite{Yuan2021} & 
		Propose an adaptive FL strategy to facilitate robust model learning from diverse edges.& Provide fast responses with accurate results while preserving users’ privacy. \\
		\hline
		Traffic Sign Recognition. & 
		\cite{Xie2022} & 
		Employ FL and spike neural networks to traffic sign recognition. & 
		Protect sensitive information and mitigating privacy risks.\\
		\hline
		Identify Vehicle Targets & 
		\cite{Xu2020} & 
		Provide a federated method to identify vehicle targets in remote sensing images. & 
		Protect privacy of remote sensing data.\\
		\hline
		Image Classification & 
		\cite{Ye2020} & 
		Propose a selective model aggregation approach toselect the most optimal local models for FL. & 
		Reduce communication costs and ensure the privacy and security of the data.\\
		\hline
		Target Recognition & 
		\cite{Zhang2021a} &
		Propose an asynchronous federated aggregation protocol to select the most optimal local models based on the quality of each model. & 
		Reduce communication costs and ensure the privacy and security of the data.\\
		\hline
		\multicolumn{4}{|c|}{In-Vehicle Driving Assistance}\\
		\hline
		Steering Angle Prediction &
		\cite{Aparna2021, Zhang2021b,Zhang2021c} & 
		Apply FL in steering angle prediction. Each vehicle acts as an edge computing device and trains local models using its own data. &  
		Reduce communication costs and ensure the privacy and security of the data.\\
		\hline
		Driver Facial Detection & 
		\cite{Zafar2021,Zhang2021d,Zhang2022b} &
		Use FL for driver status monitoring, only local in-vehicle training data is utilized. & Privacy-sensitive images of the driver do not leave the vehicle.\\
		\hline
	\end{tabular}
	\label{Tab:Federated Learning in Traffic Target Recognition}
\end{table*}

\subsection{Federated Learning in Road Target Recognition}


In the context of generic road target recognition, concerns regarding sensitive image data leakage \cite{Chen2021} and the high overhead of real-time data transmission \cite{Jallepalli2021} have prompted the use of FL for localized vehicle training. Road target recognition scenarios primarily involve target recognition and target detection, including pothole detection and traffic sign recognition.



Roads constitute a critical infrastructure for any country, and inadequate road conditions may result in vehicle damage and hazardous driving conditions. Therefore, the assessment and maintenance of road defects represent a fundamental component of ITS. Potholes, as shown in Fig.~\ref{Fig4.3_Pothole}, commonly arise due to gradual road degradation induced by various factors such as harsh weather, heavy traffic loads, and natural wear and tear of road materials. Potholes present a significant safety hazard to both vehicles and their occupants, as well as adversely impacting vehicle performance, resulting in potential damage to tires, wheels, and suspension systems. Automated road defect detection through the utilization of computer vision technology represents a promising approach to maintaining road safety and efficiency. In the context of pothole detection, Rahman \emph{et al.} \cite{Rahman2021} applied a prototype of FL to address the limitations of previous methods that were both time-consuming and prone to errors. Specifically, when a vehicle encounters a pothole, a notification response is transmitted to the central server. The central server then collates and aggregates the data obtained from various vehicles and updates the road condition periodically. Alshammari \emph{et al.} \cite{Alshammari2022} also proposed an approach to facilitate pothole detection on portable devices. Specifically, they designed a 3D pothole detection system based on FL. The proposed approach employs the YOLO model for road defect detection via size estimation, identifying objects such as patched potholes and fake road bumps. For road damage detection, Yuan \emph{et al.} \cite{Yuan2021} proposed an adaptive FL strategy for facilitate robust model learning from diverse edges.

As vehicles drive on roads, sensors and cameras play a critical role in detecting and recognizing traffic signs. Traffic signs are essential for ensuring an orderly and safe driving experience. For instance, warning signs remind drivers to be cautious  and avoid obstacles, while indicator signs help ensure  adherence to road directions. Additionally, traffic light signs aid in the controlled movement of vehicles at intersections. As such, the accurate identification and effective utilization of traffic signs have significant implications for driver assistance systems, as well as autonomous driving, enhancing  the overall safety and efficiency of vehicular transportation. Xie \emph{et al.} \cite{Xie2022} addressed  privacy in traffic sign data, which contains a substantial amount of location privacy information, by employing a prototype FL approach. The authors proposed a novel solution that integrates spike neural networks and FL to enhance  traffic sign recognition privacy. This approach holds significant promise for protecting sensitive information and mitigating privacy risks associated with traffic sign data. In the context of remote sensing image-based vehicle target recognition, Xu \emph{et al.} \cite{Xu2020} also leveraged FL to address concerns regarding the privacy of remote sensing data. This approach offers a promising solution to mitigate the risk of data leakage while also addressing issues associated with inaccurate training results and slow training speeds when dealing with single-node remote sensing data. Image classification tasks may encounter challenges associated with the diversity of image quality and computational power across client vehicles, which can have an impact on the accuracy and efficiency of FL models. To address this issue, Ye \emph{et al.} \cite{Ye2020} proposed a selective model aggregation approach to identify and select the  optimal local models for FL, thus improving overall model performance. This approach holds great potential for enhancing the accuracy and efficiency of FL in image classification tasks, particularly in situations where computational resources and image quality may vary widely among client devices. Similarly, to protect user data privacy and improve model training efficiency in the generic target recognition scenario, Zhang \emph{et al.} \cite{Zhang2021a} proposed an asynchronous federated aggregation protocol. The protocol selects the  optimal local models  based on the local quality of each classification tree.

\subsection{Federated Learning in In-Vehicle Driving Detection}


FL has shown great potential for improving in-vehicle driving detection by addressing data privacy and computational challenges in machine learning tasks. Angle prediction, crucial for autonomous driving, requires machine learning models trained on vision-based datasets. However, transmitting raw sensor data, such as that captured by cameras and lidars, can be extremely bandwidth-intensive, imposing a significant traffic load on the wireless communication system. To reduce the central computational load  and save bandwidth, FL has been applied in steering angle prediction \cite{Aparna2021, Zhang2021b}. In this approach, each vehicle acts as an edge computing device and trains local models with  data collected by its own vision sensors. The central server then updates the global model by aggregating local model parameters. Finally, the global model is shared with the vehicles for automatic steering control. This approach not only reduces the computational and communication costs, but also ensures data privacy and security. Zhang \emph{et al.} \cite{Zhang2021c} further explored model aggregation protocols. They noted that synchronous aggregation protocols are inflexible and cannot adapt to dynamic environments and heterogeneous hardware. To overcome this, they proposed an asynchronous model aggregation protocol with a sliding training window. This approach can reduce communication overhead and accelerate model training while maintaining high accuracy.


Driver facial detection is also critical for ensuring road safety by preventing accidents caused by drowsiness or distraction. However, collection and sharing  driver facial data raises privacy concerns. FL has emerged as a promising solution to address these concerns by enabling distributed model training without sharing raw data. FL can also help overcome the challenge of insufficient driver facial data by aggregating it  from multiple industry entities, such as automotive manufacturers and ride-sharing services. These entities can collaborate in a privacy-preserving manner to improve driver facial detection accuracy, ultimately enhancing road safety. In the cited papers \cite{Zafar2021,Zhang2021d}, FL was used for driver status monitoring, with the main advantage being that only local in-vehicle training data is utilized, ensuring that privacy-sensitive images of the driver do not leave the vehicle. Furthermore, the authors in \cite{Zhang2022b} proposed a pre-learning mechanism of migration learning to improve the performance of the FL system for driver drowsiness detection. Furthermore, to enhance the security of the FL system, they selected a homomorphic encryption scheme with excellent computational speed. This approach can effectively protect sensitive driver data and provide more robust security guarantees. The results demonstrate that the proposed mechanism can significantly improve the accuracy and robustness of the FL system for driver drowsiness detection.


In conclusion, the utilization of FL in autonomous driving presents an innovative solution to address the challenges of limited data volume, communication efficiency, and data privacy protection. By enabling distributed model training among multiple vehicles or devices, FL enables local data processing  while preserving user privacy. FL has shown promising results in various autonomous driving applications, including road target recognition and in-vehicle driving assistance. Further research in FL is expected to contribute significantly to the development of ITS and bring us closer to the realization of safer and more efficient autonomous driving.

\section{Federated Learning in Vehicular Edge Computing}

\subsection{Vehicular Edge Computing}


The current research proposes an intelligent-driven vehicular edge computing architecture to address the contradiction between limited network resources and massive user demands in the automotive environment \cite{Meneguette2021}. By applying mobile edge computing to the connected vehicle, vehicular edge computing enables the decentralization of computing and storage capacities. This approach can provide and manage computational resources closer to vehicles and end-users, greatly relieving the network bandwidth pressure and providing lower latency service access. Therefore, vehicular edge computing, based on the motivation and foundation of edge computing, is a promising technology to support ITS \cite{Liu2021,Hammoud2020}.


\begin{figure}[htbp]
	\centering
	\includegraphics[width=3.3in]{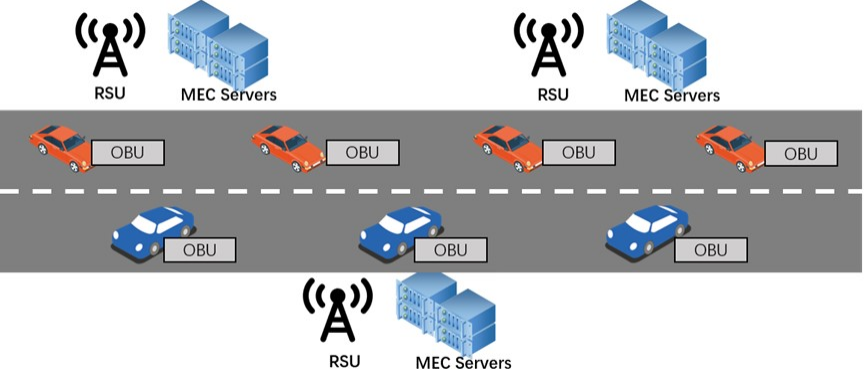}
	\caption{Framework of VEC.} 
	\label{Fig5.1_VEC}
\end{figure}

Autonomous driving systems are supported by the edge side, which comprises the On Board Unit (OBU), RSU, and Mobile Edge Computing (MEC) servers, as shown in Fig.~\ref{Fig5.1_VEC}. The OBU is responsible for environmental awareness, decision planning, and vehicle control, while the RSU provides the OBU with detailed information about the road and pedestrians. The MEC server is positioned close to the local device to reduce the processing load of the system and mitigate the latency of data transmission, ultimately enhancing the overall efficiency of the autonomous driving system.


Collaborative perception is a prominent application of edge computing in  autonomous driving. It allows vehicles to obtain sensor information from other edge nodes, thereby expanding their perception range. However, as autonomous driving levels increase and the number of smart sensors equipped on vehicles grows, an enormous amount of raw data is generated  daily. This raw data needs to be processed, fused, and feature-extracted locally and in real-time. Given that most of these tasks need to be performed in-vehicle to ensure real-time processing and response, a robust and reliable edge computing platform is essential. As such, vehicular edge computing has emerged as a new research area of focus for autonomous driving systems. In the future, it is anticipated that more than 60\% of data and applications will be generated and processed at the edge.

Vehicular edge computing has the following characteristics:

\begin{enumerate}
%

	
	\item Low latency. By bringing computing resources closer to vehicles, vehicular edge computing reduces data transmission latency, enabling real-time responses.
	
	\item Decentralized computing and storage. Vehicular edge computing allows for the decentralization of computing and storage capacity by applying mobile edge computing to the connected vehicle.
	
	\item High bandwidth. Vehicular edge computing can provide high-speed, low-latency access to data and computing resources, enabling the processing of large amounts of data generated by autonomous vehicles.
	
	
	\item Privacy and security. Vehicular edge computing can protect the privacy and security of user data by keeping the data within the vehicle or at the edge, reducing the risk of data breaches or unauthorized access.
	
\end{enumerate}



Combining edge computing with FL represents a type of vehicular edge computing. In this methodology, edge devices utilize their own collected data to train their local DL models and upload only updated models to the central server \cite{McMahan2017}. This technique helps to minimize privacy and security risks by limiting the training data to the device side only. By utilizing FL, vehicular edge computing can support collaborative learning while ensuring  data privacy and security, thereby improving the accuracy of the trained models without compromising user privacy \cite{Du2020,Lim2020}. FL in vehicular edge computing is categorized into three main types: communication efficient improved FL, resource optimized FL, and security enhanced FL. Each type is designed to address specific challenges associated with FL in vehicular edge computing. In the following, we will elaborate on these categories.

\begin{table*}[!t]
	\renewcommand{\arraystretch}{2.5}
	\caption{Federated Learning in Vehicular Edge Computing. } 
	\centering
	\begin{tabular}{|p{6cm}|p{4cm}|p{4cm}|}
		\hline 
		Applications & Key contributions Reference & Reference
		\\
		\hline
		Communication Efficient Improved FL & Reduce communication volume & 
		Model Compression  \cite{Barbieri2022,Shen2021,Li2022a} \\
		\cline{3-3}
		& & Client Selection  \cite{Ye2020a,Xiao2021,Liang2022,Liu2021a,Bao2021}	
		\\
		\cline{2-3}
		& Reduce communication rounds & \cite{Chen2021a,Liu2021a} \\
		\cline{2-3}
		& Accelerate model update &  \cite{Liang2022,Liu2021a} \\
		\hline
		
		Resource Optimized FL & Structural Heterogeneity & \cite{Xiao2021,Lu2020,Zhou2021,Danquah2021,Taik2022}\\
		\cline{2-3}
		& Fault Tolerance & \cite{Ji2021} \\
		\hline
		Security Enhanced FL & Malicious Devices Detection& \cite{Li2021,Ghimire2021,Lv2022}\\
		& Privacy Preservation & \cite{Lv2022,Li2020,Lu2019a,Kong2021}\\
		
		\hline 

	\end{tabular}
	\label{Tab:Federated Learning in Vehicular Edge Computing}
\end{table*}

\subsection{Communication Efficient Improved Federated Learning}


The communication overhead between edge devices and the central server has become a growing concern in FL, particularly in vehicular environments with limited network resources. This is primarily due to the large number of devices that send their local updates to a central server, which leads to communication bandwidth becoming a major bottleneck. To address this issue, reducing communication overhead has become a top priority. Various efforts are underway to improve communication efficiency, including reducing communication volume, minimizing the number of communication rounds, and accelerating model updates. These strategies are essential to handle the massive amount of data generated and ensure that FL can cope with the explosive growth of data in vehicular environments.

\subsubsection{Reducing Communication Volume}


In order to reduce communication overhead in FL for vehicular edge computing, several methods have been proposed in recent literature. One of these methods is model compression, which involves reducing the amount  of data transmitted in both upstream and downstream. In \cite{Barbieri2022}, a method was proposed in which vehicles optimize their local ML model and then select a fixed percentage of randomly selected parameters to share with their neighbors in each training round. The use of quantization in the ML model further saves communication resources. Similarly, Shen \emph{et al.} \cite{Shen2021} proposed a local model training algorithm based on ternary quantization, which optimizes the quantization values of parameters for different local models. This strategy  reduces the complexity of local model training and improves the overheads of upstream and downstream communication. Li \emph{et al.} \cite{Li2022a} executed optimization on  both client and server sides. On the client side, they proposed a minimal squared quantization error quantizer design, while on the server side, they proposed a closed-form solution for the optimal aggregation weights assignment. This solution minimizes the weighted sum of squared quantization errors over all active clients. These methods demonstrate the potential of model compression and quantization techniques in reducing communication overhead in FL for vehicular edge computing.


Reducing the number of clients involved in the aggregation process is another promising strategy to alleviate communication overhead in FL. The traditional federated average algorithm is a synchronous update algorithm, where all clients need to upload their model gradient information to update the server model simultaneously. As a result, the server side needs to receive most or all of the model data from the clients in each round of model aggregation, which greatly increases the data communication pressure on the server side. To address this challenge, various studies have proposed diverse client selection strategies. For example, in \cite{Ye2020a}, the authors proposed a selective model aggregation method that selects models by evaluating their local image quality and computational power and sends them to the central server. Similarly, in \cite{Xiao2021}, the optimal vehicle to participate in the learning task was selected based on position and speed. In \cite{Liang2022}, appropriate nodes were selected to participate in the aggregation process by computing their capacity, network capacity, and learning value of training samples. In \cite{Liu2021a}, clients with large amounts of data were selected to participate in the aggregation phase. Finally, in \cite{Bao2021}, client selection takes into account vehicle velocity, vehicle distribution, and wireless link connectivity between vehicles. These studies show that careful client selection can significantly reduce communication overhead in FL.

In summary, communication overhead in FL can be reduced  through model compression and client selection. Model compression can be accomplished through various techniques such as quantization and ternary quantization. The client selection optimizes the participation in the learning process based on their computational power, network capacity, and the learning value of training samples. By applying these techniques, the communication resources in FL can be saved, enabling FL to be more flexible and scalable in vehicular edge computing environments.

\subsubsection{Reducing Communication Rounds}


The communication overhead between the server and clients in FL is a critical challenge that needs to be addressed to ensure efficient data exchange. In this context, reducing the number of interactions between edges and the edge server is crucial to minimize upload times and increase download speeds. To this end, a federated stochastic variance reduced gradient-based scheme was  proposed in \cite{Chen2021a} to decrease the total number of interactions while ensuring the required accuracy. The proposed scheme optimizes the variance reduced stochastic gradient descent algorithm by employing a diagonal preconditioner and a thresholding operation to guarantee convergence. In \cite{Liu2021a}, the authors aimed to strike a balance between local computation and communication overheads by customizing local training strategies for different clients. The approach allows varying local training epochs for clients, thereby improving the trade-off between local computation and communication overheads. Specifically, clients with high computation power are allowed to perform more local epochs, while clients with low computation power are permitted to perform fewer local epochs to meet the global accuracy requirement.

These studies demonstrate that reducing communication overhead in FL can be achieved by minimizing the communication rounds between the server and clients. By adopting these techniques, the number of interactions rounds between users and the edge server can be significantly reduced while still ensuring the desired level of accuracy in vehicular edge computing environments.

\subsubsection{Accelerating Model Update}


To reduce communication overhead in FL, accelerating model updates can be another important approach. In this regard, Liang \emph{et al.} proposed a dynamic aggregation scheme in \cite{Liang2022}, which combines the advantages of both synchronous and asynchronous methods to perform semi-synchronous aggregation. The scheme adjusts the maximum server wait time dynamically based on the number of participating nodes in each round, allowing as many nodes as possible to participate in the aggregation process and hence reducing communication costs. Similarly, in \cite{Liu2021a}, the authors proposed a flexible aggregation policy that drops clients who exceed the time limitation to dynamically adjust the number of clients during the aggregation phase, thus reducing communication overheads. 

These techniques can help vehicular edge computing environments to achieve faster and more efficient model updates, significantly reduce communication overhead and enable FL to be more scalable and efficient in vehicular edge computing environments.

\subsection{Resource Optimized Federated Learning}



FL is a promising approach for collaborative machine learning that enables edge devices to train models in a decentralized manner. However, the heterogeneity of edge devices, including differences in computational power and network environment, can lead to imbalanced training times and inefficient global model aggregation. Straggler devices with weaker computational power may significantly delay the model aggregation, while devices with unstable networks may drop out of the training process, compromising the efficiency of FL. To address these issues, appropriate resource allocation and fault tolerance mechanisms need to be considered to minimize training time and improve FL's efficiency. Several studies have proposed different optimization methods for formulating the FL process.

\subsubsection{Structural Heterogeneity}

Various studies  have developed  algorithms for efficient resource allocation in FL that account for the heterogeneous nature of edge devices. For instance, in \cite{Xiao2021}, the authors considered the specific characteristics of vehicular edge computing environments, such as vehicle position and velocity, to formulate a min-max optimization problem. The objective was to jointly optimize the onboard computation capability, transmission power, and local model accuracy to achieve the minimum cost in the worst-case scenario of FL. By considering the worst-case scenario, this approach ensures robustness and reliability in FL training, even under challenging network conditions. In \cite{Lu2020}, the authors proposed an asynchronous FL scheme that aims to improve the efficiency of FL by selecting the participating nodes to minimize the total cost. This approach considers the computation, communication, and storage costs of the edge devices, and dynamically selects a subset of participating nodes in each round of FL based on their cost-efficiency scores.  Similarly, in \cite{Zhou2021}, a unified data and resource management framework was proposed to cluster vehicles on the road based on their mobility and resource characteristics. This framework partitions computationally intensive tasks' data and assigns them to individual vehicles in the cluster for parallel execution.

Additionally, some studies have proposed clustering clients. For example, in \cite{Danquah2021}, a data and resource management framework was proposed that organizes vehicles on the road into clusters based on their mobility and resource characteristics, partitioning the computationally intensive tasks among the individual vehicles in the cluster for parallel execution. Furthermore, in \cite{Taik2022}, the authors designed an architecture and corresponding FL process for clustered FL in vehicular environments, aiming to enhance the scalability of FL. They formulated a joint cluster-head selection and resource block allocation problem taking into account mobility and data properties.


\subsubsection{Fault Tolerance}

To address the issue of stragglers in FL, Ji \emph{et al.} proposed an edge-assisted FL approach that leverages edge computing to alleviate the computational burdens for straggler devices. The proposed approach allows stragglers to offload partial computation to the edge server and utilizes the server’s computing power to assist clients in model training \cite{Ji2021}. This approach not only reduces the time needed for stragglers to complete their tasks but also enhances the overall efficiency of the FL process.

By addressing the challenges of heterogeneous edge devices, appropriate resource allocation and fault tolerance mechanisms can significantly improve the efficiency of FL. These optimization methods can minimize training time and reduce energy consumption. Leveraging such  approaches, FL has the potential to unlock the full potential of vehicular edge computing and pave the way for ITS.

\subsection{Security Enhanced Federated Learning}


\subsubsection{Malicious Devices Detection}
In FL, edge devices process their local data without sending it to the central server, thus preserving user privacy. However, since the server has limited control over the edge devices, it cannot fully trust them. Some edge devices may have malicious intent and falsify or poison their training data, resulting in useless or even harmful model updates. Therefore, protecting the integrity and authenticity of the training data is a critical issue in FL. Several studies have proposed solutions to address this problem, including data poisoning detection, outlier detection, and secure aggregation techniques, which can help mitigate the impact of malicious edge devices on the FL process. These solutions enable the server to verify the authenticity and quality of the data provided by edge devices and ensure the overall reliability and security of the FL process.


To mitigate the risk of malicious devices participating in FL, various methods have been proposed. In \cite{Li2021}, the authors addressed the issue of malicious mobile-edge computing servers and vehicles in FL by adopting blockchain technology to incentivize vehicles to contribute and reduce the influence of malicious participants. They also proposed a traceable identity-based privacy-preserving scheme to protect vehicular message privacy. In another study, Ghimire \emph{et al.} \cite{Ghimire2021} applied the quickest change detection technique to detect changes in the statistical properties of the model parameters sent by participating devices. This approach facilitates the identification of malicious clients. In \cite{Lv2022}, the authors proposed a blockchain-based FL scheme to detect misbehavior by coordinating multiple distributed edge devices while ensuring data security and privacy. These approaches aim to address the challenge of trustworthiness in FL by detecting and mitigating the influence of malicious devices on the training process.


\subsubsection{Privacy Preservation}
The adoption of FL offers a promising solution to mitigate the privacy leakage concerns. However, the challenge of indirect privacy leakages still remains  significant. These indirect leakages can potentially compromise the privacy of user data, thereby making it imperative to address them effectively.



To enhance the privacy preservation of clients, differential privacy serves as a key strategy of protecting  sensitive data. When querying data from a database, differential privacy reduces the chances of records being identified while maximizing query accuracy by introducing noise to the raw data \cite{Li2020}. In \cite{Lu2019a}, the authors incorporated local differential privacy into FL to protect the privacy of updated local models. Similarly, in \cite{Lv2022}, differential privacy with the Gaussian mechanism was leveraged to provide strict privacy protection for the model on the blockchain.

Homomorphic encryption is another privacy strategy that is often applied in FL to prevent information leakage during the exchange of parameters between clients. It is an encryption mechanism that encodes parameters prior to an addition or multiplication operation, resulting in an equivalent result to a non-encoded function \cite{Li2020}. In \cite{Kong2021}, the authors proposed a privacy-preserving model aggregation scheme that uses a homomorphic threshold cryptosystem for key establishment and updates. The scheme leverages the homomorphic properties of the threshold cryptosystems to allow clients to perform secure addition and multiplication operations on the encrypted model updates, while ensuring the final aggregated model remains encrypted.

In conclusion, security and privacy are critical challenges in FL. To mitigate these issues, a variety of methods, including blockchain technology and the quickest change detection technique, have been developed to detect malicious devices and prevent attacks. Furthermore, by incorporating differential privacy and homomorphic encryption, it is possible to further protect the privacy of clients' data and model parameters exchanged between clients. These security-enhancing techniques can increase the reliability and trustworthiness of FL, making it a promising approach for collaborative learning while ensuring privacy and security of the data. 

\section{Conclusion}


The integration of FL into ITS applications provides a collaborative training approach that protects data privacy. By  performing distributed model training on local devices, FL ensures that sensitive user data is never leaked to other organizations or servers, thus satisfying the need for data privacy protection. This  survey provides a comprehensive review of the current research hotspots in combining FL and ITS, including traffic flow prediction, traffic target recognition and vehicular edge computing, which can serve as a valuable resource for future work in this field. The application of FL in ITS has yielded promising results and has the potential to bring significant benefits. Our review highlights the potential of FL as a powerful tool for achieving sustainable and harmonious urban transportation systems.

\bibliographystyle{ieeetr}

\bibliography{refs}

\end{document}